\documentclass{sig-alternate}

\usepackage{acronym}
\usepackage{subfigure}
\usepackage{algorithm}
\usepackage{algcompatible}
\usepackage{color}
\usepackage{url}

\acrodef{MIP}{Mixed Integer Program\-ming}
\acrodef{P2P}{Peer-to-Peer}
\acrodef{MOG}{Multiplayer Online Game}
\acrodef{WLAN}{Wireless Local Area Network}
\acrodef{MANET}{Mobile Ad-Hoc Network}
\acrodef{AI}{Artificial Intel\-li\-gen\-ce}

\makeatletter
\def\@copyrightspace{\relax}
\makeatother

\begin{document}

\title{Mobile Online Gaming via Resource Sharing\footnotemark}

\def\sharedaffiliation{%
\end{tabular}
\begin{tabular}{c}}

\numberofauthors{2}

\author{
\alignauthor Stefano Ferretti\\
\email{sferrett@cs.unibo.it}
\alignauthor Gabriele D'Angelo\\
\email{g.dangelo@unibo.it}
\sharedaffiliation
\affaddr{Department of Computer Science}\\
\affaddr{University of Bologna}\\
\affaddr{Mura A. Zamboni 7, I-40127 Bologna, Italy}\\
}

\maketitle

\footnotetext{The publisher version of this paper is available at \url{http://dx.doi.org/10.4108/icst.simutools.2012.247720}.
\textbf{{\color{red}Please cite this paper as: Stefano Ferretti, Gabriele D'Angelo. Mobile Online Gaming via Resource Sharing. Proceedings of 3nd ICST/CREATE-NET Workshop on DIstributed SImulation and Online gaming (DISIO 2012). In conjunction with SIMUTools 2012. Desenzano, Italy, March 2012. ISBN: 978-1-936968-47-3.}}}

\abstract{Mobile gaming presents a number of main issues which remain open. 
These are concerned mainly with connectivity, computational capacities,
memory and battery constraints.
In this paper, we discuss the design of a fully distributed approach for the support of mobile Multiplayer Online Games (MOGs). In mobile environments, several features might be exploited to enable resource sharing among multiple devices / game consoles owned by different mobile users. We show the advantages of trading computing / networking facilities among mobile players. 
This operation mode opens a wide number of interesting sharing scenarios, thus promoting the deployment of novel mobile online games. In particular, once mobile nodes make their resource available for the community, 
it becomes possible to distribute the software modules that compose the game engine. This allows to distribute the workload for the game advancement management.
We claim that resource sharing is in unison with the idea of ludic activity that is behind MOGs. Hence, such schemes can be profitably employed in these contexts.
}

\terms{Algorithms, Performance, Theory}


\section{Introduction}

In the last years, \ac{MOG} technologies have been widely studied. Several solutions have been identified which are concerned with many aspects of \acp{MOG}. An open aspect relates to the development of mobile games. Nowadays, portable game consoles exist which are quite powerful computing devices, equipped with networking technologies. For this reason, from a certain point of view the player does not expect great differences between playing using a mobile game console and using a PC (apart from the obvious graphic limitations and the small screen of a mobile device). He/she expects to be able to generate inputs, perceive the evolution of the game in real-time, interact with other distributed players in the (more or less) same way when he/she uses a traditional PC/game console.
However, the truth is that several technical constraints limit the gaming activities of mobile players.

Mobile devices are primarily constrained by the limited battery capacity. Every task, computation or communication has a power cost; hence these activities should be limited as much as possible in order to preserve the battery lifetime. Communications pass through wireless channels, which can be less reliable and offer lower data rates than traditional wired ones. This should be took into consideration during the design of the game. In particular, different alternatives are possible depending whether the game is played among mobile players located in the same (limited) geographical area rather than when distributed players and nodes connected through the Internet must be reached.
In the first case, in fact, a \ac{MANET} can be built to optimize the game update distribution. In the second case, these forms of interaction are usually based on mobile devices that connect to their nearest access point to access the Internet. However, the proliferation of heterogeneous devices with different capabilities gives rise to new scenarios that promote the cooperation among individuals in order to guarantee the provision of ``always on'' services~\cite{iwcmc}.

Resource sharing and optimization opens novel interesting scenarios for the deployment of mobile \acp{MOG}. There are two ways to optimize the digital resources usage.
First, mobile users have many different portable devices in their pockets
and suitcase, each of them with specific hardware and software
characteristics. Quite often such devices are not enabled for seamless
interaction with other devices belonging to the same owner. Instead, the use of the networking capabilities of a user mobile phone might, for instance, relieve the gaming console from performing long-range communications with other nodes participating to the game. Hence, the game console might communicate with the user's mobile phone through a short range networking technology (e.g.~Bluetooth, ZigBee); in turn, the mobile phone relays messages generated by (or directed to) the game console by employing its long-range network technology (e.g.~3G, Wi-Fi).

Then, there is the opportunity of sharing resources among different people. The possibility for a user to exploit, in a~\ac{P2P} and altruistic way, computing facilities owned by (known and trusted) neighbor players requires mechanisms for automatic negotiation~\cite{iwcmc}. In this case, a node might share its computation with its neighbor, for instance when it updates the game state due to novel received game events. Alternatively, a node might act as a relay for other ones during the distribution of game updates.

In a recent paper, we have identified mechanisms to promote resource sharing among distributed nodes in wireless communication environments~\cite{iwcmc}. In this paper, we describe the main possibilities arising from resource sharing for the deployment of novel mobile \acp{MOG}. Specifically, we review the main software components composing an online game, and discuss how and when these software modules can be distributed and/or replicated at mobile nodes.

In fact, certain software modules have to be executed at all nodes, such as those for managing inputs from the player, rendering the game state evolution and performing basic networking to enable communication with other nodes. Other software modules, instead, can be distributed (and replicated) at different nodes such as those for the game state management, the network overlay management (when the communication occurs through a \ac{MANET}), and several sub-components of the game engine, such as those for simulating the physics of game objects, identifying collisions, etc. 
Usually, such modules are implemented as event-driven software components. This eases the distribution of these modules, since the game engine becomes a discrete-event based distributed system.

It is clear that when a node executes a task, additional communication is required to share the outputs of this module with other nodes. This has an additional cost, besides that required for executing the software module itself. Hence, in certain cases these software modules might be replicated to diminish the communication workload of the nodes with other ones and also to improve scalability and fault-tolerance, thus preventing that a single node becomes the bottleneck of the system. 
When certain modules are replicated, synchronization and consistency management techniques must be employed \cite{debs,ieee_tmm}.

We claim that resource sharing is in unison with the idea of ludic activity that is behind \ac{MOG}s. Hence, such schemes can be profitably employed in these contexts.

The rest of this paper is organized as follows. Section \ref{sec:archi} discusses the main architectural solutions which may be employed to support \acp{MOG}, with specific attention to games deployed for mobile devices. Section \ref{sec:proposal} discusses how effective mobile architectures for the support of mobile \acp{MOG} may be devised, which resort to resource sharing and to the distribution of software modules. Section \ref{sec:modules} outlines some main aspects to consider when allocating and distributing software modules in a mobile \ac{MOG} architecture. Finally, Section \ref{sec:conc} provides some concluding remarks.

\section{Wireless MOG Architectures}
\label{sec:archi}

This section presents an overview of the distributed software architectures that can be employed to build \acp{MOG} on wireless networks. Basically, a first distinction can be made between those gaming applications that involve mobile users placed in a localized geographical area only, rather than those games whose nodes need to communicate through the Internet. 

When players are all confined in the same \ac{WLAN}, their mobile nodes might organize in an ad-hoc manner, i.e.~they form a \ac{MANET}, and each node reaches another node using a multi-hop relays. Hence, all the communications remain within the \ac{MANET}.

On the opposite case, nodes communicate through the Internet using some structured communication architecture such as, for instance, a 3G cellular networking technology (e.g.~UMTS), rather than some structured Wi-Fi network.
At the data-link layer, each node communicates with its access point directly, as in all traditional wireless Internet-based communications.

There is however a hybrid solution, according to which mobile nodes on the same geographical area communicate in an ad-hoc manner; but when needed, data can pass through the Internet, usually to reach some server and/or other distributed nodes not belonging to the \ac{MANET}.
In such a case, one or more (possibly all) nodes in the \ac{MANET} must be able to send messages outside the local network. Hence, in this case some specific wireless communication technology must be exploited which is different to the network technology employed to interact in an ad-hoc fashion with other mobile neighbor nodes. In other words, some mobile node might exploit multiple network interfaces concurrently.

Despite the underlying communication architecture, at the application layer different solutions are possible, ranging from the client/server to the \ac{P2P} scheme~\cite{ace}. Our proposed solution adopts a hybrid approach. Next sections are devoted to describe this solution in detail.

Note that in this section, we consider each node is composed of a unique device. Hence, we do not take into account that a player might have actually different devices that might be coordinated. If this is the case, then the interaction of this set of devices might be properly configured, so that the mobile node forms a sort of ``digital organism'' able to exploit all its devices effectively~\cite{iwcmc}.

\subsection{The Client/Server Scheme}

\begin{figure*}[t]
\centering%
\subfigure[\label{fig:cs-no_overlay}]{\includegraphics[width=.48\linewidth]{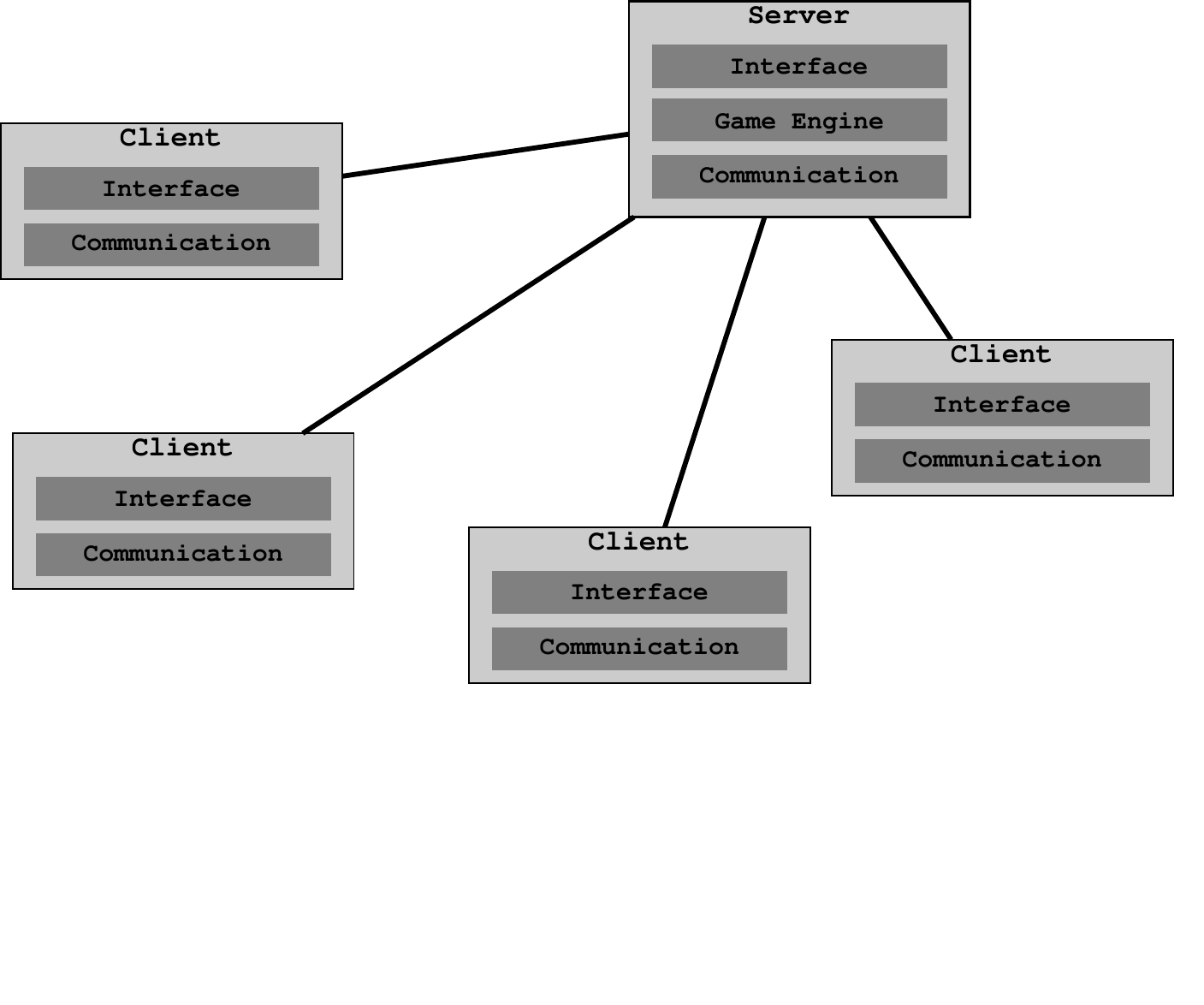}}%
\subfigure[\label{fig:cs-overlay}]{\includegraphics[width=.48\linewidth]{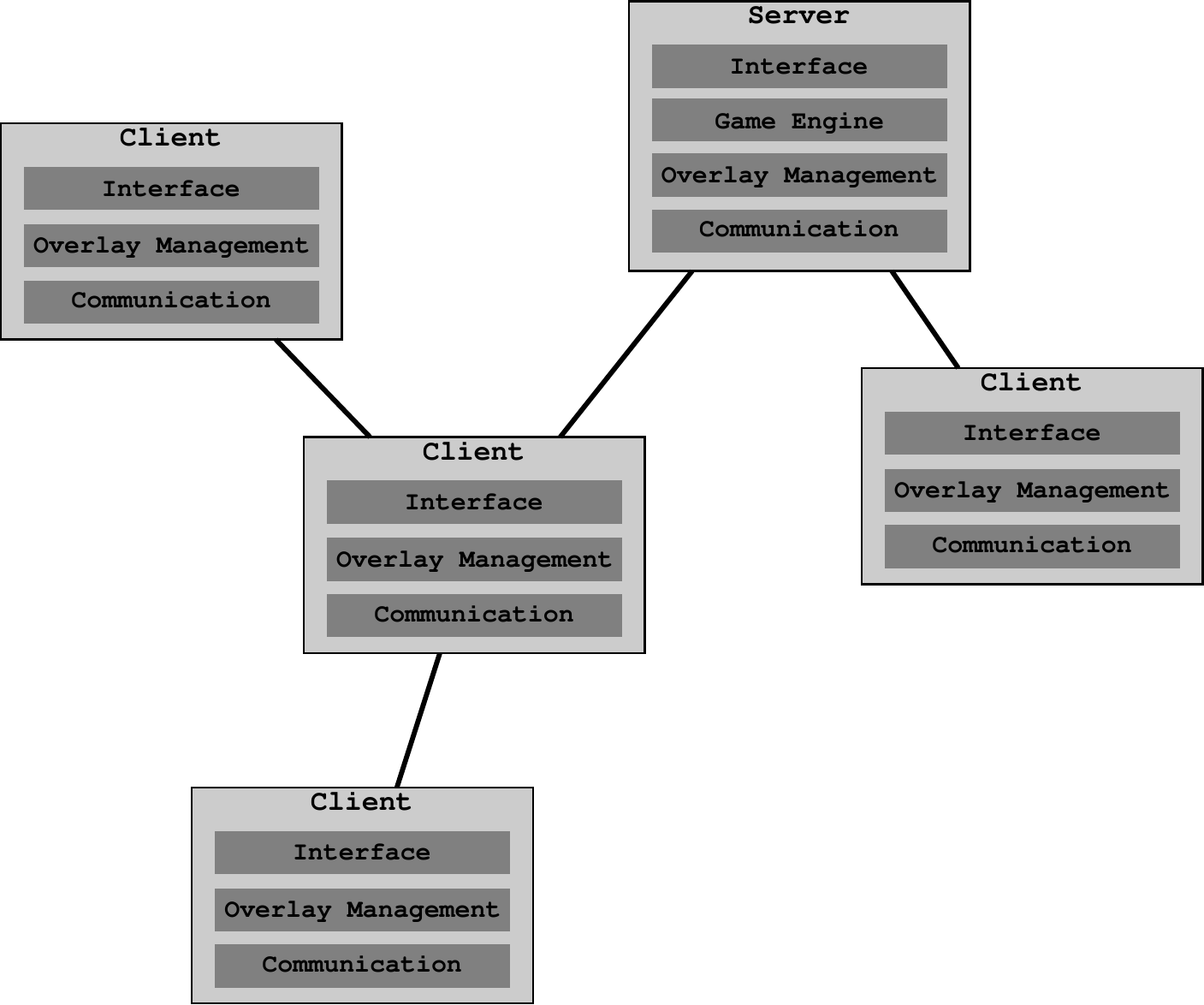}}
\caption{Client-Server Model: a) all clients connected to the server; b) with overlay}
\label{fig:cs}
\end{figure*}

The client/server scheme is widely employed in traditional \acp{MOG}, where a fixed node is usually employed as the central node that maintains the game state and interacts with all other nodes in the network. It has been recognized that such a solution is quite simple to implement but, on the other hand, it raises several reliability and fault-tolerance issues. 

In few words, client nodes execute the following two software components: i) the user interface that collects inputs from the user and to render the game state evolution; ii) a software module in charge of managing the network communication.
The server has additional software modules to manage the game state evolution (we neglect additional software components that are used to perform offline operations, e.g.~accounting). 

Figure~\ref{fig:cs} shows such an architectural approach. While in common scenarios the server is in direct connection with all the clients (Figure~\ref{fig:cs-no_overlay}), when a mesh overlay structure is employed (like  in a \ac{MANET}) messages from certain client nodes might require multiple hops before reaching the server (Figure~\ref{fig:cs-overlay}). Note that in this case each node must run an ``overlayManagement'' software module in charge of managing the overlay and control the routing of messages passing through that node.
This approach presents several disadvantages when employed over wireless networks. Indeed, the communication among the server and clients might be quite unstable. Nodes may move during their interactions; this would require reconfigurations of the overlay, due to the fact that some communication links may become unavailable.
There are problems related to the fact that mobile nodes have restricted battery power. When a node fails, also the communications passing through it fails consequently. In substance, the overlay management may require an intense configuration and communication overhead.


\subsection{The Peer-to-Peer Scheme}

\begin{figure*}[t]
\centering%
\includegraphics[width=.7\linewidth]{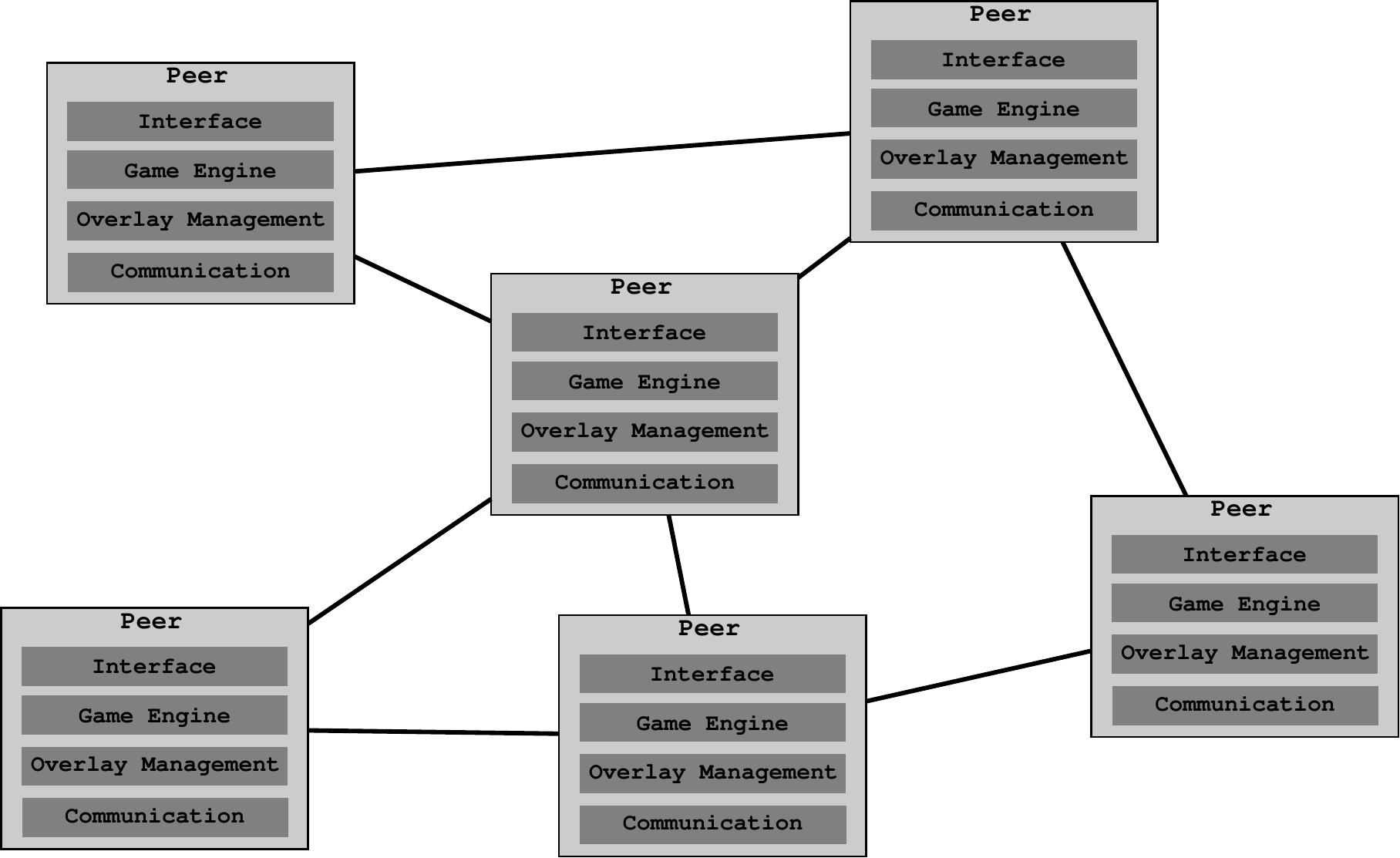}%
\caption{Peer-to-Peer Model}
\label{fig:p2p}
\end{figure*}

In a pure \ac{P2P} system, each mobile node stores and manages its own copy of the game state. Hence, each peer executes all the software modules which in the client/server approach were executed on the server only (see Figure~\ref{fig:p2p}).

As concerns the overlay topology, a possibility is to organize nodes based on a fully connected network. Actually, this approach is impractical (or even impossible) in certain \ac{MANET}s, since a node might not be able to reach all the nodes with a single-hop message transmission.
The adoption of a mesh overlay network for disseminating game events represents a viable and more scalable choice. Of course, a careful management of the overlay is needed, so as to ensure that game events are delivered to all nodes in a timely fashion and that the game advances in real-time.

A possible technical solution is to resort to broadcast (or multicast) wireless communications in \acp{MANET}. The idea is that a single wireless message can reach multiple mobile nodes; this can be reasonably exploited in \acp{MOG}, since usually all game updates must be sent to all (or a subset of) peers. Such broadcast schemes can be employed with multi-hop communication protocols, in order to span the message through the entire overlay.


The clear advantage of a \ac{P2P} scheme is that it removes the presence of a single server, which represents the bottleneck and single point of failure of the system. However, every node must execute all software modules, hence augmenting the computational work at each node (this has consequences on the battery consumption also). Moreover, since a \ac{P2P} architecture is structured typically as a mesh overlay, churns must be viably managed.


\subsection{Game Software Modules}

In the figures above, for the sake of a simpler presentation we referred to the game software modules by taking into consideration a coarse range of generic functionalities such as the ``interface'', ``game engine'', ``overlay management'', ``communication''.
Actually, a more detailed set of software modules can be listed, based on the tasks associated. Based on their functionalities, some of them can be assigned to different nodes to be executed in a distributed architecture. 
\begin{itemize}
\item \textbf{Input management}.\\
  This software module includes all the functionalities needed to manage game events produced by the player, which causes a game state update.
\item \textbf{Audio management and sound system}.\\
  This module is responsible to manage all the audio sounds to be played, associated to particular game events occurring during the game. It is also responsible for the background audio, which is usually played during a game session. Mixing is required among audio played in background and sounds associated to specific game events.
\item \textbf{Scene graph engine}.\\
  Here, we include all the functionalities needed to render the game on the player screen. This includes management of (2D/3D) objects, skeletal animation, texturing and imaging, lighting and shading, rendering of the terrain, water, smoke, clouds and the like.
\item \textbf{Physics system}.\\
  This module includes point and rigid body dynamics, soft body dynamics, fluid dynamics and kinematics in general. In substance, this is a dynamics simulation component, which is responsible for managing and solving the simulated physical forces affecting the simulated game objects.
\item \textbf{Collision detection}.\\
  This is the module in charge for determining when two game objects collide during their movement. For the purpose of reducing the computation needs, usually simplified objects are employed; then, these simplified meshes (bounding boxes, spheres, convex hulls) are used for determining collisions. Besides these mentioned simplifications on the game objects' representations, such a task is computation demanding, nevertheless.
  Such calculations must be performed periodically, at a constant frame-rate. The higher the frame-rate the more accurate the model for determining collisions. Hence, this software modules should be executed on computationally efficient nodes.
\item \textbf{Game state management}.\\
  This module maintains the state of all the game objects (characters and other virtual objects) on the game map and, based on the events produced by all participants, updates it. It checks the validity of game events produced by players and works in strict collaboration with the collision detection module and the physics system. Indeed, in certain cases we might think that the two modules mentioned here above are included in this one, being their tasks employed for computing game advancements.
\item \textbf{Virtual map and scene storing service}.\\
  Usually, the information that describes the virtual map area is replicated on each host. However, especially when the game has a huge virtual map, then each node can maintain only the part of the virtual world where its character is located at that time; as soon as the character moves on another region, then the novel virtual area is downloaded on the terminal. If this is the case, then a server (or a set of servers) is responsible to maintain the complete virtual world; this is necessary to ensure that the node can download the virtual world description when it needs it. When a distributed \ac{P2P} architecture is employed, a server node might still be employed that maintains the whole virtual map, rather than distributing the map among different nodes. 
  Of course, parts of the map might be replicated for the purposes of reliability and of an easier and more efficient data distribution.
\item \textbf{Artificial intelligence}.\\
  This software module is responsible for the management of virtual bots interacting with users in the virtual world. Some node executes the \ac{AI} module that decides the moves each bot performs during the game evolution. Usually, all these tasks are in charge to the server. When a decentralized solution is employed, the management of different virtual bots can be delegated to different nodes, such as in a mobile multi-agent system.
\item \textbf{Finite state machine management}.\\
  This module governs the evolution of the game as determined by users' actions. Based on the actual game state and on the events generated by users, the game can evolve based on certain rules. All this can be implemented through a finite state machine. (Actually, finite state machines can be employed also to specify the behavior of virtual bots; thus, they represent a possible tool employed to realize the \ac{AI} of virtual players.)
\item \textbf{Prediction schemes and dead reckoning}.\\
  Dead reckoning is a technique employed in \acp{MOG} to reduce the effects of network induced delays and losses by applying prediction schemes~\cite{Pantel:2002}. Each node routinely uses dead reckoning to predict where an actor might be located at a given time, based on past information on its last known kinematic state. When correctly employed, it allows to avoid that each node sends game state updates that can be easily inferred from previous information. By resorting to such an approach, the use of the network is reduced and the game advancement fastened. 
  The quality of the prediction is thus quite important in order to ensure that all players perceive the game evolution in a consistent way.
\item \textbf{Accounting and Score Management}.\\
  These functionalities, concerned with the management of accounts and scores of players participating to the game, are prone to cheating. Hence, when executed on a \ac{P2P} architecture, proper strategies must be took into account so as to prevent that some malicious actor alters some information, or acquires data it is not allowed to access~\cite{Chambers:2005,Ferretti08}.
\item \textbf{Networking}.\\
  Common communication capabilities are required to let mobile nodes to communicate. Depending on the game implementation both UDP or TCP transport level protocols can be employed. Indeed, while UDP is recognized as the typical choice for transporting data of real-time multimedia applications, some works suggest that some sort of tuned TCP represents an interesting alternative~\cite{Griwodz:2006,ieee_tmm}. While in typical situations the mobile node is configured to exploit a single network interface card, in general situations it might be the case when the mobile node is enabled to concurrently exploit multiple network interface cards. This would promote interesting novel communication scenarios; for instance, it would allow a node to interact with other nodes geographically located near it through some short range communication technology, while using at the same time long range communication technologies to reach other hosts on the Internet. Not only, sophisticated communication schemes can be employed to let the node to exploit different network interface cards to communicate with another host, in order to guarantee seamless interactions with it \cite{jss}.
\item \textbf{Overlay management}.\\
  We already mentioned that in a \ac{MANET}, nodes operate as both end hosts and routers, forwarding packets wirelessly towards other mobile nodes that may not be within the direct transmission range of each other~\cite{castro}. Thus, routing strategies are needed which are able to adapt depending on the availability and position of nodes. Each node must thus maintain a table with its neighbors, in order to relay messages. Moreover, reconfiguration strategies are needed to adapt to the overlay changes.
\end{itemize}

\section{On the Optimal Organization of Resources}
\label{sec:proposal}

\begin{figure*}[t]
\centering%
\includegraphics[width=.7\linewidth]{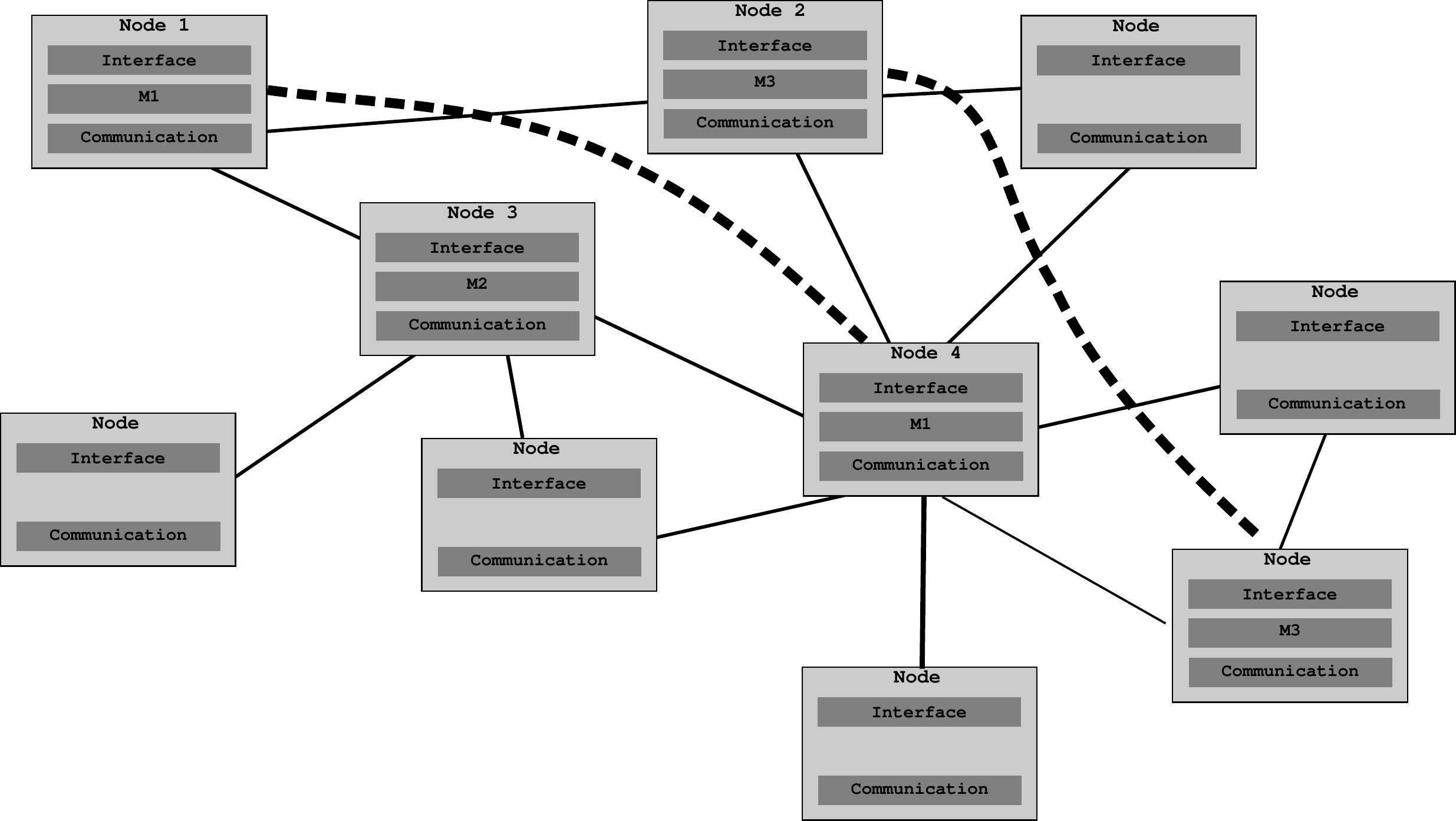}%
\caption{Distribution of Software Modules on the Mesh}
\label{fig:archi}
\end{figure*}

In this section, we propose a distributed solution for the deployment of ubiquitous gaming applications. All devices available to each player are dynamically and adaptively configured depending on: 
\begin{enumerate}
  \item the devices themselves, 
  \item the environment in which they are deployed, and 
  \item the computational and communication capabilities of devices of other players. 
\end{enumerate}

The architecture must thus provide configuration protocols for the intra Personal Area Networking (PAN), to automatically organize all devices belonging to a single player. At the same time, it is necessary to identify algorithms and mechanisms for the simultaneous and adaptive use of different communication networks in an opportunistic fashion. In fact, the overall goal is to optimize the interactions across all players in the wireless overlay.
Finally, we need mechanisms for the efficient distribution of software modules composing the game engine. This allows to create a distributed game engine whose execution spans through the whole mobile \ac{MOG} architecture.

In essence, the idea is to resort to a hybrid architecture, where each software module can be executed on a specific node, depending on its computational and communication capacities.
Hence, computation is distributed among different nodes. Similarly to a client-server approach, these services (e.g.~computation of the game state evolution) must be made available to other nodes that do not have that software module locally active.
Software modules might be replicated, so as to augment reliability and distribute the communication load with other nodes that require the service.


When needed, also the game engine can be divided into subcomponents, so as to distribute the computational load even further among players. For instance, while a node might be selected to perform collision detection, another node might be in charge of performing the AI of some virtual bots, and so on. (This specific example might lead to several security concerns, and in this case cheating prevention schemes would be required.)
Each node must run the interaction and communication modules, so as to permit interactions with the user and other nodes, respectively.

Several heuristics and optimization strategies can be implemented to distribute and replicate all the software modules and the virtual world map. These schemes must take into account:
\begin{itemize}
 \item the geographical location of players, and thus the overlay topology that may be built, based on players' position;
 \item the node capabilities in terms of computation capacity, communication, memory, status of the battery;
 \item the need to interact with nodes external to the \ac{MANET}, i.e.~Internet nodes. This depends on the specific game application.
\end{itemize}


To make this possible, we need to manage and optimize i) the interactions among the different devices that each single player has in his/her hands while playing and, ii) the interactions among different players in the overlay mesh.
Then, it is possible to distribute all the software modules and create a smart game management architecture that may improve the quality of the gaming experience to all the mobile players.

\subsection{Optimizing the Player's Devices}

Full interaction among all digital devices hold by each single player might promote the deployment of effective mobile \acp{MOG}~\cite{iwcmc}. This implies to optimize the use of available networking technologies, such as short-range communication technologies, e.g.~Bluetooth, infrared, ZigBee, etc., for communicating with other players which are located near the considered node. 
The goal is to find the best configuration for all the devices in use at each player. Specifically, based on the computational capacities of each device within the PAN, the battery levels, and the available network interfaces, devices must be configured so as to identify a primary computation entity, a primary gateway to send/receive data from the outside world, secondary network interfaces (e.g.~short range ones) to allow communications with neighbor players.

According to our architecture, all devices exchange their profiles among each other, in order to enable a proper system configuration. Different alternatives exist to characterize profiles of devices, such as, for instance, CC/PP \cite{CCPP}. Such information is exploited to identify the coordinator, i.e.~the device that acts as the resource manager of the player's devices. To accomplish this task, all devices' profiles must be distributed among the whole local device set, and some distributed algorithm must be executed to elect the coordinator. A similar approach must be employed to identify which device is to act as the primary gateway that manages communications with the outside world.

Upon a proper organization of the user's devices, that device set might be seen as a unique computation/communi\-ca\-tion node with a set of features which is composed by the aggregation of featured of single devices.

\subsection{Optimizing Interactions Among Nodes}

Peer players must be provided with a set of protocols to interact with others within their \ac{MANET}. These protocols would allow a peer to opportunistically and dynamically adapt the interaction with its neighboring peers, by selecting the best communication protocol among the available ones (e.g. Always Best Connected, ABC) \cite{aict,jss}. The identification of the best available network may be based on different criteria such as bandwidth, connection cost, battery level and so on. Any of these criteria, alone or together with the others, can be used for assessing the best available network at any time.


\section{Distribution of Software Mo\-du\-les over the Mesh}
\label{sec:modules}

Figure~\ref{fig:archi} shows an example of a mobile \ac{MOG} architecture, where nodes are connected through a mesh overlay (small, continuous lines). Every node executes the software module interface, that refers to the capabilities of managing inputs from the user and outputs to be shown to the user; moreover, every node executes the communication module, which allows to interact with other nodes. As already mentioned, these two modules are mandatory to every node.

Then, other modules are distributed to be executed on certain nodes on the overlay only. Since the figure refers to an hypothetical example, we avoided to list specific names of modules composing the game engine. Hence, in the figure we used generic names $M1, M2, \ldots$ Some of these modules are replicated (see larger, dashed lines in the figure). In this case, other nodes select to which node refer for that service. For instance, node $1$ and node $4$ both execute module $M1$. Hence, others select the one that requires less distance hops. 

In substance, when configuring such an architecture, there are some main aspects to take into consideration, i.e.~allocation of software modules, generation of the overlay mesh, distribution of nodes acting as clients for a given service executed at another node, synchronization of states managed at replicated services so as to ensure state consistency.

\subsection{Modules Distribution}

Here, we discuss which software modules can be distributed and or replicated, among those mentioned in the previous section.
\begin{itemize}
\item \textbf{Input management, Audio management and sound system, Scene graph engine}.\\
  As already outlined above, and shown in Figure~\ref{fig:archi}, these modules must be executed on each node.
\item \textbf{Physics system, Collision detection, Game state management}.\\
  These are main modules that can be distributed on some nodes. They can be replicated also, for the sake of scalability and fault tolerance. However, the outputs of each of these modules are important for other modules. In substance, these modules form a sort of core game engine. Hence, probably the best choice is to execute them on the same nodes.
\item \textbf{Virtual map and scene storing service}.\\
  As already mentioned, this module (and mostly, the data composing the game virtual map) can be assigned to a third server rather than being distributed on different nodes.
\item \textbf{Artificial intelligence}.\\
  Each virtual bot and its \ac{AI} can be executed on a given peer node, and possibly different bots might be managed at different peers.
\item \textbf{Finite state machine management}.\\
  Being one of the main functionalities of a server (in a client/server architecture), this module can be distributed.
\item \textbf{Prediction schemes and dead reckoning}.\\
  This module is client-specific, executed to hide communication latencies, hence it must be executed on every node.
\item \textbf{Accounting and Score Management}.\\
  It can be assigned to a distributed (trusted) node.
\item \textbf{Networking}.\\
  It must be executed on each node.
\item \textbf{Overlay management}.\\
  It can be distributed on nodes that would be in charge of deciding how to organize the overlay.
\end{itemize}

\subsection{Allocation of Software Modules}

When distributing software modules, several mechanisms can be employed to have a fair allocation over distributed nodes, taking into account the computation and communication capacities of the nodes. One possible solution is to employ simple heuristics. For instance, having an estimation of the workload of a software components, which depends on the number of users to be served, given the number of nodes it is possible to identify the number of replicated nodes acting as ``servers'' for that service. Then, such distribution can be performed by ranking nodes based on their computation/communication/battery capacities and on their geographical location (probably, it would be preferable to uniformly distributing the services on the overlay).
This task should be repeated for each service.

Another option is to optimize the allocation by matching demands for executing different services and offers, using some kind of market-based approach in which requests are handled
through ascending clock auctions~\cite{Ausubel2004,HuangHCP07,Kelly04,stokely}. Such an approach can be performed by running a distributed algorithm to carry on the auction; then, the resource allocation problem can be treated as a classic optimization problem, which is used to compute the maximum number of allocations that can be matched.

It is clear that when several software modules are distributed over different nodes, there are several security concerns that must be considered. Cheating is a main problem in \ac{P2P} \acp{MOG}; hence viable strategies must be enforced to prevent and detect cheats~\cite{Ferretti08}.
Moreover, some form of authentication and authorization must be considered in order to identify players that ask services to other distributed nodes.

\subsection{Management of the Overlay Mesh}

The task of generating and managing the overlay mesh can be performed by resorting to one among the plethora of proposals that manage mesh overlays in a \ac{MANET}. Examples are works presented in~\cite{Canourgues,castro,Li08,Wongsaardsakul}.

\subsection{Associating Clients to Software Modules}

Once software modules have been distributed and a mesh overlay has been built, nodes not running a given module must ask for updates to a node running the service. Also in this case, several options are possible, each one with its pros and cons. Each node might ask to its nearer node running the service. Thus, upon a novel update to be disseminated, that node would receive the update in the fewest number of hops. However, several updates from different services travel through a path; moreover, a given node might be overloaded with too many nodes to serve. These issues might influence the performance of this approach.
Moreover, since an update must be multicast to several nodes, effective strategies might be employed that ensure that the mesh overlay is covered in a minimal number of hops.

In substance, the problem here refers to building a publish-subscribe scheme over a mobile overlay. Works that deal with this issues have been presented in~\cite{Baehni:2005,Anceaume:2002,Huang:2004,Rezende:2008}.

\subsection{Synchronization}

Replicating software modules at distributed nodes provide several advantages. Indeed, this approach augments the scalability of the system, its fault-tolerance and might improve the responsiveness of the interaction between the client node and that running the service, due to the fact that this node has to manage a lower number of client nodes.

However, when the replicated software module deals with the game state management, then synchronization algorithms must be executed between nodes running the service. This is important to guarantee that each node perceives a consistent evolution of the game state. There are different alternatives in this case, such as~\cite{Cronin:2004,debs,mauve,ieee_tmm}.

%
%

\section{Conclusions}\label{sec:conc}

Mobile \ac{MOG} architectures may benefit from viable resource sharing strategies.
In this paper, we have discussed a methodology to optimize the interactions of mobile players into dynamic and heterogeneous environments. The idea is to optimize the use and interaction of the devices available to each player, through dynamic and adaptive configuration strategies (optimization of the PAN). Then, interactions among players can be optimized using both the available communication infrastructures and P2P ad-hoc interactions.
Based on these resource sharing mechanisms, software modules composing the game engine can be distributed (and replicated) across the whole gaming network. This would distribute the workload needed for the game advancement, hence resulting in an improved resource usage in a mobile environment.

As concerns the general deployment of the proposed scheme in a real
distributed system, there are some open problems that require further
investigation. Security issues are particularly important: for
instance, cheating is a primary issue to deal with in this case. Moreover, authorization and authentication must be enforced to verify the identity of users that exploit resources of their neighbor players.

\bibliographystyle{abbrv}
\bibliography{ecosystem}

\end{document}